\documentclass[10pt]{article}
\usepackage{amsmath}
\usepackage{amssymb}
\usepackage{graphicx}
\usepackage{cite}
\usepackage{color} 
\usepackage[labelfont=bf,labelsep=period,justification=raggedright]{caption}
\makeatletter
\renewcommand{\@biblabel}[1]{\quad#1.}
\makeatother

\date{}

\begin{document}

\begin{flushleft}
{\Large
\textbf{Synchronization efficiency in coupled stochastic oscillators: \\ The role of connection topology}
}
\bigskip
\\
G. Reenaroy Devi$^1$, R. K. Brojen Singh$^{2*}$, Ram Ramaswamy$^{2,3}$
\\
\bigskip
$^1$ Centre for Interdisciplinary Research in Basic Sciences, Jamia Millia Islamia, New Delhi 110025, India.\\
$^2$School of Computational and Integrative Sciences, Jawaharlal Nehru University, New Delhi-110067, India.\\ $^3$School of Physical Sciences, Jawaharlal Nehru University, New Delhi-110067, India.
\\
\bigskip
$\ast$ Corresponding author, E-mail:  R.K. Brojen Singh - brojen@jnu.ac.in, \\

\end{flushleft}

\section*{Abstract}
We study the efficiency of synchronization in ensembles of identical coupled stochastic oscillator systems. By deriving a chemical Langevin equation, we measure the rate at which the  systems synchronize. The rate at which the difference in the Hilbert phases of the systemsevolve provides a suitable order parameter, and a 2--dimensional recurrence plot further facilitates the analysis of stochastic synchrony. We find that a global mean--field coupling effects the most rapid approach to global synchrony, and that when the number of ``information carrying'' molecular species increases, the rate of synchrony increases. The Langevin analysis is complemented by numerical simulations.
\bigskip

{\bf Keywords:} Cell signaling, synchronization, diffusive coupling, mean-field coupling, Neurospora crassa.

\subsection*{Introduction}

The nature of synchrony in stochastic dynamical systems has been a subject of interest given the ubiquity of both stochasticity and synchrony (or more generally, strong temporal correlativity) in a variety of natural systems \cite{glas,pik,ram}. The dynamics of biological systems, particularly at the cellular or subcellular level, are known to be subject to large fluctuations, and it is clear that diverse processes need to be fine--tuned temporally in order that any biological ``system'' is able to function. Other examples can be found in areas ranging from neuroscience to the nature of financial markets, and indeed, in other instances when a systems approach is applicable \cite{imb,ran,boc}. 

The coupling of autonomous dynamical systems can be effected through a variety of different mechanisms such as subjecting them to a common driving signal \cite{pec}, or through diffusion \cite{koc,nan,che} or through a mean-field \cite{pik1,nan}. One of the most widely studied effects of such coupling is the emergence of new collective behaviour, for instance synchronization \cite{pik,ram}.    As has been established through extensive work in the past decades, synchronization comes in a number of flavours: phase \cite{ros1}, diffusive \cite{fis}, frequency \cite{asi}, delay \cite{men}, complete \cite{pec}, generalized \cite{rul} synchronization, based on which variables are being synchronized and how the synchrony is manifest. Different measures or order parameters to gauge the rate of synchrony under the various scenarios have been devised:  for instance the phase difference $\Delta\phi$ as a function of time \cite{ros1}, Lyapunov exponents \cite{pik1}, permutation entropy \cite{ban}, phase locking value \cite{lac} etc.  Several of these scenarios apply in the presence of noise, both external and internal \cite{ban,liu,ros}. 

In the past few years, the study of stochastic synchronization, namely the emergence of temporal correlations between stochastic dynamical systems has been a major area of study \cite{ram}.  The approach to phase synchrony in such cases depends both on the nature of the coupling as well on details relating to time delay or coupling topology,  and this issue forms the central question of the present paper: How does synchronization efficiency depend on these different factors? 

This is a central issue in the related context of how the dynamics in a group of systems become correlated. To give a biological instance, circadian rhythms in bacterial organisms such as {\it Neurospora crassa} are globally correlated through environmental fluctuations. A number of biological rhythms are  based on genetic processes that originate in negative feedback circuits \cite{har,dun,you,rep} to generate a periodic expression of specific genes within a cell \cite{wy,dun}. Clock proteins like {\it FRQ}~in {\it N. crassa}, or  {\it PER}~and {\it TIM}~in {\it D. melanogaster} are responsible for the genetic regulation of clock genes \cite{gol,gol1}. Environmental fluctuations are known to be a means of correlating  a group of cells or unicellular bacteria, most notably via the mechanism of quorum sensing \cite{bas}. Synchrony appears to be essential for information processing \cite{glas} and is  the consequence of  cell--to--cell communication via specific coupling mechanisms \cite{aih}. 

In order to focus on the measurement of the rate of synchrony for various coupling mechanisms and  to identify the factors on which this rate depends, we study coupled stochastic oscillators within the  ``chemical Langevin'' formalism.  This is presented in Section II for different coupling mechanisms. In Section III we examine a circadian oscillator model which we study in detail both numerically and analytically. Our results are summarized in Section IV.

\subsection*{Chemical Langevin equation formulism of coupled oscillators}

The random interaction of molecules in a well stirred mesoscopic system leads the dynamics of the variables in the system to noise-driven stochastic process \cite{gil1,gil2,gil3,gil4,kam}. Consider the state of the system at any instant of time $t$ is defined by a configurational state vector, $\mathcal{C}(t)$=$[X_1(t),$$X_2(t)$,...,$X_N(t)]^{T}$, where $N$ distinct molecular species are interacting via $M$ elementary reaction channels in the system of the following type,
\begin{eqnarray}
\alpha_{1\mu}X_{1}+\cdots \alpha_{N\mu}X_{N}\stackrel{k_{\mu}}{\rightarrow}\beta_{1\mu}X_{1}+\cdots \beta_{N\mu}X_{N}
\label{eq1}
\end{eqnarray}
where, $\mu$ is the reaction number index: $\mu=1,...,N$, $\alpha_{i\mu}$ and $\beta_{i\mu}$ are co-efficients to define stoichiometric matrix $\nu_{i\mu}=(\alpha_{i\mu}-\beta_{i\mu})$, and $k_{\mu}$ is the $\mu$th macroscopic rate of reaction. The system evolves with various random reaction fired at random interval of time with decay or/and creation of molecular species at any reaction event \cite{gil1,gil2} which leads to the change in configurational state of the system. This allows to define a configurational probability $P(\mathcal{C};t)$ as the probability to get this state change in an interval of time $[t,t+\Delta t]$. Then the time evolution of configurational probability $P(\mathcal{C};t)$ obeys Chemical Master equation (CME) \cite{mcq,gil1}. The CME in general provides detail mesoscopic description of chemical kinetics, but it is very difficult to solve for complex systems \cite{gil2}. 

The chemical Langevin equation (CLE) formalism is one method to approximate CME to simpler continuous Markov type equations by keeping conditions which are applicable in natural systems \cite{gil3}, and the accuracy of this CLE is found to be more than those of other formalisms such as linear noise approximation \cite{gri}. The approximation can be done by allowing to define a function $A(\mathcal{C},\Delta t)$ as the number of a particular reaction fired during an interval of time $[t,t+\Delta t]$ with $\Delta t\rangle 0$. This is followed by excellent approximations by imposing two conditions, firstly, imposing small $\Delta t$ limit such that the values of propensity functions $\omega(\mathcal{C})$ of the reactions remain constant during $[t,t+\Delta t]$, and secondly imposing large $\Delta t$ limit which in turn leads to $\omega(\mathcal{C}(t))\Delta t\rangle\rangle 1$. These two conditions allow $A$ to approximate to statistically independent Poisson random variable and then the Poisson random variable is replaced by normal variable with the same mean and variance. Both the conditions are true in natural practice for large population limit. Then linearizing the normal variable, and defining macroscopic molecular concentration vector $\{\bf x\}=\frac{1}{V}\mathcal{C}(t)$, where $V$ is the systems size, we have general CLE,
\begin{eqnarray}
\label{lan1}
\frac{d{\bf x(t)}}{dt}=F[\omega({\bf x(t)}),{\bf \nu}]+G[\omega({\bf x(t)}),{\bf \nu},{\bf \xi},V]
\end{eqnarray}
where, $F=\sum^{M}_{i=1}\nu_{ij}\omega_i[x(t)]$ is the macroscopic contribution term and $G=\frac{1}{\sqrt{V}}\sum^{M}_{i=1}\nu_{ij}\left[\omega_i\{x(t)\}\right]^{1/2}$ is the stochastic contribution term to the dynamics. $\xi_i$ = $lim_{dt\rightarrow 0}N_i(0,1)/\sqrt{dt}$ is uncorrelated, statistically independent random noise parameters which satisfy $\xi_i(t)\xi_j(t^\prime)$ = $\delta_{ij}\delta(t-t^\prime)$. 

Now consider two identical interacting stochastic systems with configurations $U(t)$=$[x_1(t),$$x_2(t)$,...,$x_N(t)]^{T}$ and $U^\prime(t)$=$[x_1^\prime(t),$$x_2^\prime(t)$,...,$x_N^\prime(t)]^{T}$, where their dynamics are given by CLE of the type equation (\ref{lan1}), $\stackrel{.}{x}=U(x)$ and $\stackrel{.}{x}^\prime=U^\prime(x^\prime)$ respectively. The interaction of the two stochastic systems can be allowed by choosing one or more "coupler" molecular species and introducing one of the various coupling mechanisms, namely, direct (master-slave) \cite{pec}, diffusive \cite{nan}, mean-field \cite{piko} coupling etc via $y$. We then construct a larger 2N-dimensional stochastic system, $H(z)$ whose dynamics is given by, ${\bf \stackrel{.}{z}}=H(z)$ where, ${\bf z=(x_1,\dots, x_N,x^\prime_1,\dots,x^\prime_N)}$. Then the system is devided into sub-systems, $U(x)$ and $U^\prime(x^\prime)$ consisting of independent reaction sets with corresponding dynamics and a third arbitrary reducible sub-system $S(y)$ in the same configurational space $H(z)$ formed by $m$ extra reaction channels introduced via $2n$ coupling molecular species $y=(x_{i+1},\dots,x_{i+n},x_{i+1}^\prime,\dots,x_{i+n}^\prime)^T$. If there is no coupling between the two systems, $S\rightarrow 0$, otherwise $S$ is finite and it depends on various factors and parameters such as directionality of coupling, coupling constants etc. This $S$ associates with signal or signals common or diffused between the two sub-systems derived from the logical operations or extra reaction channels responsible for the coupling. This leads to the following dynamics of the $S$, $U$ and $U^\prime$ sub-systems given by, 
\begin{eqnarray}
\label{lan2}
\frac{d{\bf y}(t)}{dt}&=&H({\bf y,\nu,\xi,V})+S_y({\bf y,V,\epsilon,\xi})\\
\frac{d{\bf x}(t)}{dt}&=&U({\bf x,\nu,\xi,V}),~~~~~({\bf x\ne y})\\
\frac{d{\bf x^\prime}(t)}{dt}&=&U^\prime({\bf x^\prime,\nu,\xi,V})~~~~~({\bf x^\prime\ne y})
\end{eqnarray}
where, $U=[H(x_1),\dots,H(x_N)]^T$ and $U^\prime=[H(x_1^\prime),\dots,H(x_N^\prime)]^T$ are of the form of equation (\ref{lan1}) for uncoupled casees. The last term $S_y({\bf y,V,\epsilon,\xi})$ is the coupling term added to the corresponding coupling variables. $\epsilon$ is the coupling constants between the two subsystems. The functional form of $S_y$ depends on how the two systems coupled and type of coupling mechanism, for example: (i) ${\bf Direct~coupling}:$ if $x_j=x_j^\prime$, $(j=i+1,\dots,i+n)$, then $S_y\rightarrow 0$ and the other variables will achieve synchronization \cite{pec}, (ii) ${\bf Diffusive~coupling}:$ if the diffusive reactions are incorporated to couple the two systems, then $S_y$ is finite with those number of diffusive reactions (uni/bi directional) to form $S_y$ and synchronization can be achieved among the rest of the variables for sufficiently large values of $\epsilon$, (iii) ${\bf Mean-field~coupling}:$ if mean information of the two systems is allowed to use for signal transduction between the two systems by constructing an arbitrary molecular species, $\eta_j=\frac{1}{R}\sum_{k=1}^{R}x_j^k$ ($R=2$)  and allowed to diffuse to interact with corresponding molecular species, $x_j$ and $x_j^\prime$ in the two sub-systems via diffusive reactions, then synchronization will be exhibited among the other variables, and so on.

We can generalize this coupling method for $L$ identical stochastic systems by constructing a large stochastic system $H=[U^{1}(N,M),U^{2}(N,M),\dots,U^{L}(N,M),S_y(M_R)]$, where $M_R$ is the total number of extra reaction channels allowed depending upon the type of coupling and the way how they couple among them. If we look for steady state solutions of CLE (3)-(5) by applying conditions, $\stackrel{.}{x}=0$, $\stackrel{.}{x}^\prime=0$ and $\stackrel{.}{y}=0$, then we obtain $S_y(M_R)=0$.

\subsubsection*{Synchronization efficiency: diffusive coupling}

We first consider diffusive coupling mechanism between two stochastic systems to understand the rate synchronization among the coupled systems. If ${\bf y}=(x_\alpha,x_\alpha^\prime)$ is taken to be couplers which can diffuse in and out of the systems (bidirectional), this coupling between any two stochastic systems $U=[x_1, x_2, ..., x_N]^{-1}$ and $U^\prime=[x_1^\prime, x_2^\prime, ..., x_N^\prime]^{-1}$ can be achieved by incorporating two extra reaction channels; $x_\alpha\stackrel{c}\rightarrow x_\alpha^\prime$; $x_\alpha^\prime\stackrel{c^\prime}\rightarrow x_\alpha$. When the coupling constants $c$ and $c^\prime$ are strong enough, the dynamics of other variables of the systems i.e. $\{(x_i,x_i^\prime); i=1,2,...,N, i\ne\alpha\}$ will achieve synchronization. In this case $H$ will have 2N variables, $2M+2$ reaction channels: $M$ reaction channels for each sub-systems $U$ and $U^\prime$, and 2 for$S_y(2)$. We take $c=c^\prime$ for simple case and the functional forms of two $S_y=(S_{x_\alpha},S_{x^\prime_\alpha})$ for two variables are given by the following CLEs derived from equation (\ref{lan1}) for the two reactions,
\begin{eqnarray}
\label{lan4}
&&S_{x_\alpha}(x_\alpha,x_\alpha^\prime,V,c,\xi)\nonumber\\
&&~~~~~=c[x_\alpha^\prime-x_\alpha]+\sqrt{\frac{c}{V}}\left[\sqrt{x_\alpha^\prime}\xi_r-\sqrt{x_\alpha}\xi_s\right]\nonumber \\
&&S_{x^\prime_\alpha}(x_\alpha^\prime,x_\alpha,V,c,\xi)\nonumber\\
&&~~~~~=c[x_\alpha-x_\alpha^\prime]+\sqrt{\frac{c}{V}}\left[\sqrt{x_\alpha}\xi_r^\prime-\sqrt{x_\alpha^\prime}\xi_s^\prime\right]
\end{eqnarray}
Substituting these functions $S_1$ and $S_2$ in equations (3)-(5) we have CLE for the two coupled stochastic systems. Now we look solution of the CLEs for stability condition. Since $\stackrel{.}{x}=U(x)$, $\stackrel{.}{x}^\prime=U^\prime(x)$ and $H(y)$ is a part of $U$ and $U^\prime$ for CLE (3)-(5) and (6), we have the stability conditions, $\stackrel{.}{x}=0$, $\stackrel{.}{x}^\prime=0$. This conditions give us $S_{x_\alpha}=0$ and $S_{x^\prime_\alpha}=0$ respectively and after solving these two equations, we get the following result.
\begin{eqnarray}
\label{cv}
\frac{1}{cV}=\Lambda_s(x_\alpha,x_\alpha^\prime)
\end{eqnarray}
where, $\Lambda_s(x_\alpha,x_\alpha^\prime)$=$\left[\frac{x_\alpha^\prime-x_\alpha}{\sqrt{x_\alpha^\prime}\xi_r-\sqrt{x_\alpha}\xi_s}\right]^2$. For the sake of simplicity we take $\xi_r\sim\xi_s\sim \sqrt{B}$ or $\xi_r^\prime\sim\xi_s^\prime\sim \sqrt{B}$, where $B$ is a constant and then taking $x_\alpha\sim x_\alpha^\prime$, we have, $\Lambda_s(x_\alpha,x_\alpha^\prime)$ $\sim$ $\frac{4x_\alpha}{B}$. So for bidirectional diffusive coupling of single molecular species, the coupling constant is related to system's size by, $c\sim \left[\frac{B}{4x_\alpha}\right]\frac{1}{V}$.

Now we consider two molecular species $x_\alpha$ and $x_\beta$ as "couplers" and are allowed to diffuse among the two systems $U$ and $U^\prime$ in the similar fashion discussed above by constructing $S_y({\bf y})=(x_\alpha,x_\beta,x_\alpha^\prime,x_\beta^\prime)^T$ by introducing four extra reaction channels: $x_\alpha\stackrel{c}\rightarrow x_\alpha^\prime$; $x_\alpha^\prime\stackrel{c^\prime}\rightarrow x_\alpha$; $x_\beta\stackrel{c^{\prime\prime}}\rightarrow x_\beta^\prime$; $x_\beta^\prime\stackrel{c^{\prime\prime\prime}}\rightarrow x_\beta$ giving $S=(S_{x_\alpha},S_{x^\prime_\alpha},S_{x_\beta},S_{x^\prime_\beta})$. We then substitute these forms of $S_y$ in the equation (3)-(5), and the equations become CLE for the two coupled stochastic systems. We then look for steady state solutions by applying steady state conditions which give all four $S=0$. Solving the equations for $x_\alpha$ and $x_\alpha^\prime$ and keeping $c=c^\prime$ we have the relation between $c$ and $V$ for $x_\alpha$ is given by equation (\ref{cv}). Similarly keeping $c^{\prime\prime}=c^{\prime\prime\prime}$, the relation between $c^{\prime\prime}$ and $V$ for $x_\beta$ can be obtained by solving the equations $S=0$ for $x_\beta$ and $x_\beta^\prime$ which leas to the following equation,
\begin{eqnarray}
\label{cvd}
\frac{1}{c^{\prime\prime} V}=\Lambda_s^\prime(x_\beta,x_\beta^\prime)
\end{eqnarray}
where, we take $\xi_d\sim\xi_e\sim \sqrt{B^\prime}$ ($\xi_d^\prime\sim\xi_e^\prime\sim \sqrt{B^\prime}$) and then taking $x_\beta\sim x_\beta^\prime$, we have, $\Lambda_s^\prime(x_\beta,x_\beta^\prime)$ $\sim$ $\frac{4x_\beta}{B^\prime}$ such that $c^{\prime\prime}\sim \left[\frac{B^\prime}{4x_\beta}\right]\frac{1}{V}$. 

The rate constant of a chemical reaction is the product of number of collisions per unit time and the probability that any given collision in the reaction takes place \cite{nak}. If the two sub-systems are coupled via diffusion of two molecular species with two different coupling constants $c$ and $c^{\prime\prime}$, the equivalent coupling constant $c^\prime$ of the two diffusing rate constants will follow "parallel resistance law" hypothesis \cite{gav1,gav2,bur} given by $\frac{1}{c^\prime}=\frac{1}{c}+\frac{1}{c^{\prime\prime}}$. The hypothesis can be generalized for $L$ different rate constants given by vector ${c}=(c_1,c_2,\dots,c_L)^T$ to obtain equivalent rate constant $\frac{1}{c}=\frac{1}{c_1}+\frac{1}{c_2}+\dots+\frac{1}{c_L}$.  From the equations (\ref{cv}) and (\ref{cvd}), the equivalent coupling constant $c^\prime$ is solved using this hypothesis,
\begin{eqnarray}
\label{cv2}
c^\prime\sim&\frac{B}{4V}\left[\frac{1}{x_\alpha+x_\beta}\right]
\end{eqnarray}
where, we take $B=B^\prime$ for the sake of simplicity and since they are random numbers.

\subsubsection*{Synchronization efficiency: mean-field coupling}

We now consider $E$ identical subsystems, and define an arbitrary hypothetical mean field molecular species, $\eta_{\alpha}(t)=\frac{1}{K}\sum_{j=1}^{K}x_\alpha^{j}(t)$, which is the average information carried by $K$ subsystems and allowed to signal transduction via molecular species $x_\alpha$ in the ensemble. Depending on the number $\Lambda$ and the topology of the ensemble of $E$ sub-systems, the mean-field coupling could be nearest neighbour ($K=$2 (1-dimension), 4(2-dimension), 6(3-dimension)), next to nearest neighbours and global ($K=E$).  Any two oscillators $U_j$ and $U_{j+1}$ in the ensemble of $E$ sub-system can be coupled via mean-field by allowing $\eta_{\alpha}^j(t)$ and $\eta_{\alpha}^{j+1}(t)$ to diffuse between the two subsystems via two extra reactions, $\eta_{\alpha}^j\stackrel{\epsilon}\rightarrow x_{\alpha}^{j+1}$; $\eta_{\alpha}^{j+1}\stackrel{\epsilon^\prime}\rightarrow x_{\alpha}^j$. Thus we can construct $S(x_{\alpha}^j,x_{\alpha}^{j+1})$ because $\eta_{\alpha}^j=\eta_{\alpha}^j(x_{\alpha}^j)$ and $\eta_{\alpha}^{j+1}=\eta_{\alpha}^{j+1}(x_{\alpha}^{j+1})$. This mechanism is similar to that of diffusive kind but exchange of mean information takes place. Proceeding in the same way as we did in diffusive coupling case, we can get the set of CLE by substituting the calculated $S$ in (3)-(5). Taking $\epsilon=\epsilon^\prime$ the relation between $\epsilon$ and $V$ is obtained as,
\begin{eqnarray}
\label{cvm}
\epsilon=\frac{1}{\Lambda_n(\eta,\eta^\prime)V}\sim\frac{1}{V}\frac{D}{4\eta}
\end{eqnarray}
where, $\Lambda_n(\eta,\eta^\prime)$=$\left[\frac{\eta^\prime-\eta}{\sqrt{\eta^\prime}\xi_g-\sqrt{\eta}\xi_h}\right]^2$ and $\xi_d\sim\xi_e\sim \sqrt{D}$ ($\xi_d^\prime\sim\xi_e^\prime\sim \sqrt{D}$). 

We then couple the two oscillators $U^j$ and $U^{j+1}$ in the ensemble with mean-field coupling mechanism via two molecular species $x_\alpha^j$ and $x_\beta^{j+1}$ by defining two corresponding hypothetical mean-field molecular species $\eta_{\alpha}(t)=\frac{1}{K}\sum_{j=1}^{K}x_\alpha^{j}(t)$ and $\psi_{\beta}(t)=\frac{1}{K}\sum_{j=1}^{K}x_\beta^{j}(t)$ respectively. These $\eta_\alpha$ and $\psi_\beta$ are allowed to diffuse in the sub-systems and interact with the local molecular species $x_\alpha$ and $x_\beta$ in the respective sub-systems. This can be done by defining four extra reactions: $\eta_{\alpha}^j\stackrel{\epsilon}\rightarrow x_{\alpha}^{j+1}$; $\eta_{\alpha}^{j+1}\stackrel{\epsilon^\prime}\rightarrow x_{\alpha}^j$; $\psi_{\beta}^j\stackrel{\epsilon^{\prime\prime}}\rightarrow x_{\beta}^{j+1}$; $\psi_{\beta}^{j+1}\stackrel{\epsilon^{\prime\prime\prime}}\rightarrow x_{\beta}^j$. This leads us to construct $S(x_{\alpha}^j,x_{\alpha}^{j+1},x_{\beta}^j,x_{\beta}^{j+1})$ because $\eta_\alpha$ and $\psi_\beta$ depend on $x_\alpha$ and $x_\beta$ respectively. Then substituting $S$ forms in equations (3)-(5) we get CLE of the coupling mechanism. Then putting $\epsilon=\epsilon^\prime$ for $\eta_\alpha$ and $\epsilon^{\prime\prime}=\epsilon^{\prime\prime\prime}$ for $\psi_\beta$, we solve the CLE for steady state conditions and using the" parallel resistance law" hypothesis for $\epsilon$ and $\epsilon^{\prime\prime}$, we get the equivalent rate constant $\epsilon^\prime$ as in the following,
\begin{eqnarray}
\label{cvm2}
\epsilon^\prime\sim\frac{D}{4V}\frac{1}{\eta+\psi} 
\end{eqnarray}
As we see from the above equation that the functional form of the rate constants in both mean field and diffusive couplings are similar.
\begin{figure}
\label{fig1}
\begin{center}
\includegraphics[height=150 pt,width=8.5cm]{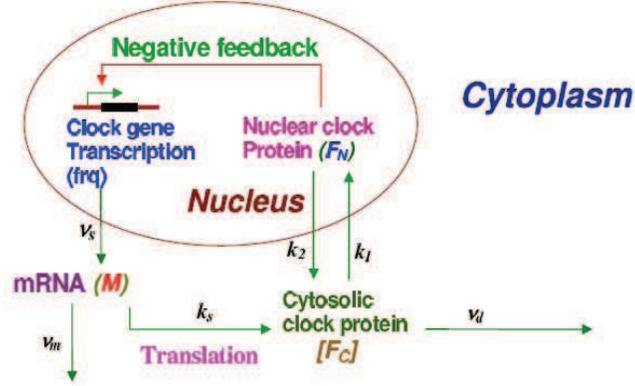}
\caption{Biochemical reaction network of circadian rhythm in N. crassa due to Gonze and Goldbeter \cite{gon2}.}
\end{center}
\end{figure}

\begin{figure}
\label{fig2}
\begin{center}
\includegraphics[height=250pt]{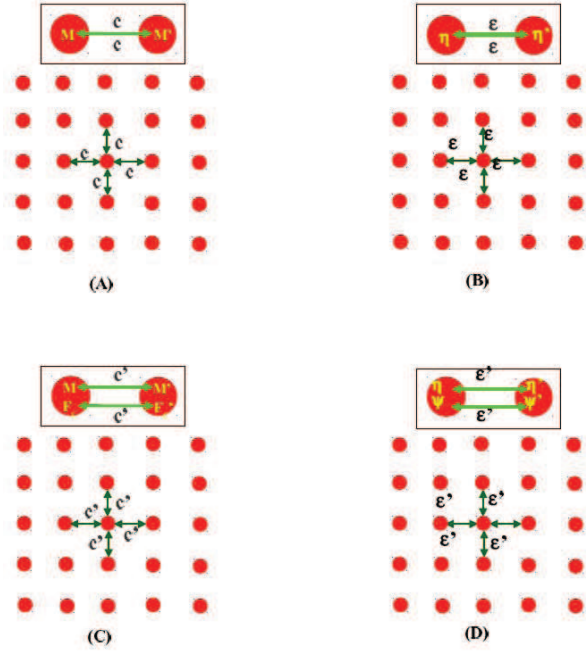}
\caption{Coupling mechanisms we considered in our simulation: (A) Nearest neighbour single molecular species ($M$ is coupling molecular species) bidirectional diffusive coupling, (B) Nearest neighbour single molecular species ($\eta=\frac{1}{N}\sum_{i=1}^NM^{[i]}$ is coupling molecular species) bidirectional mean-field coupling, (C) Nearest neighbour double molecular species ($M$ and $F_C$ are coupling molecular species) bidirectional diffusive coupling and (D) Nearest neighbour double molecular species ($\eta=\frac{1}{N}\sum_{i=1}^NM^{[i]}$ and $\psi=\frac{1}{N}\sum_{i=1}^NF_C^{[i]}$ are coupling molecular species) bidirectional mean-field coupling.}
\end{center}
\end{figure}
\begin{figure}
\label{fig3}
\begin{center}
\includegraphics[height=170pt]{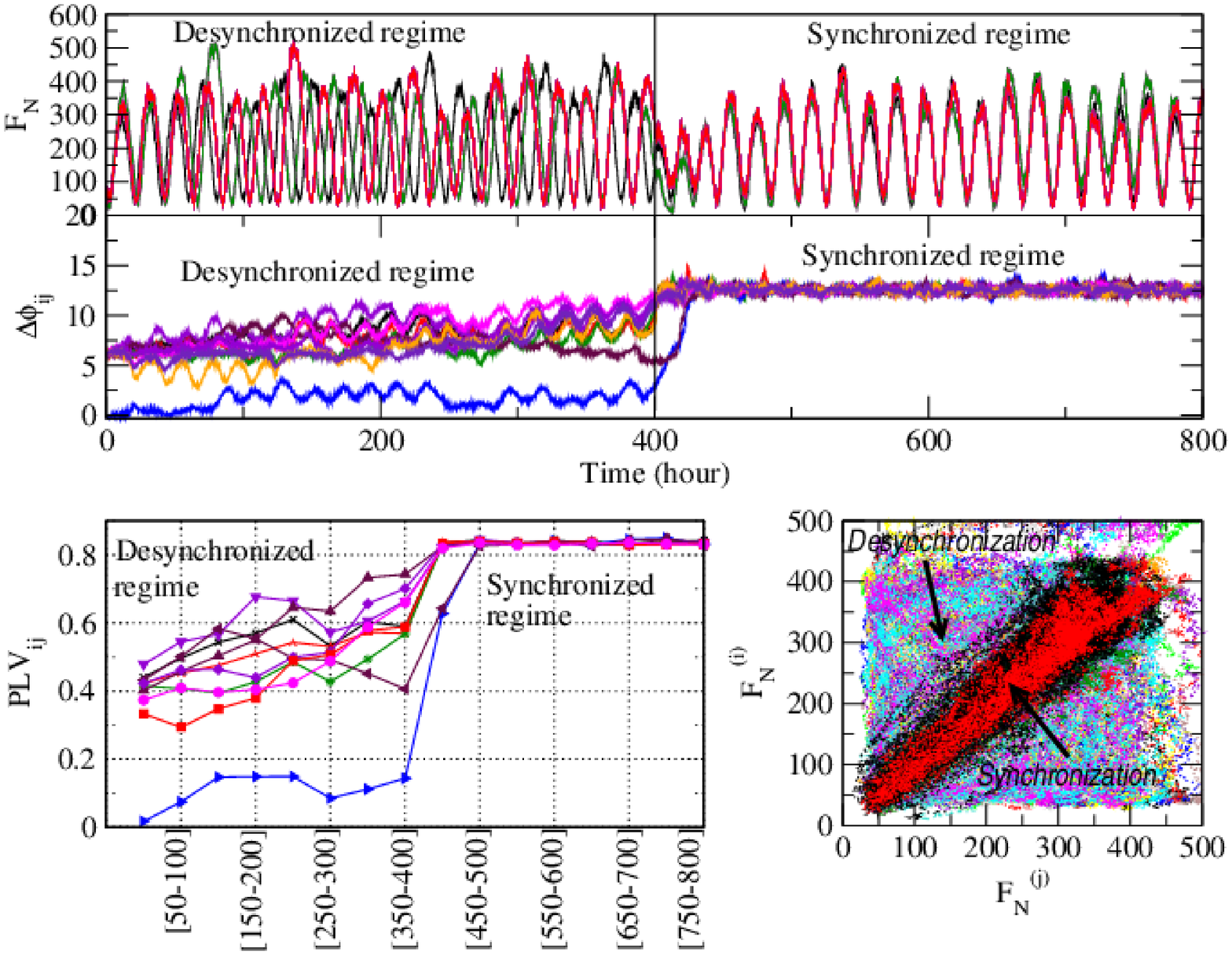}
\caption{Plots of nuclear protein $F_N(t)$ for 10 oscillators out of 50 oscillators as a function of time, for $V=100$ and $\epsilon=1.4$ showing synchronized and desynchronized regimes. Phase differences of pairs of oscillators, $\Delta\phi_{ij}$, $i,j=$ $\{$(1,2), (1,3),(1,4), (1,5), (1,6), (1,7), (1,8), (1,9), (1,10), (1,50)$\}$ as a function of time i.e. phase plot, phase lock values of the corresponding set pairs of oscillators and recurrence plot are also shown in second, third and fourth panels respectively.}
\end{center}
\end{figure}

\subsubsection*{Comparison of the coupling mechanisms}

The comparison of the coupling constants in all the four coupling mechanisms is done so that one can compare the rate of synchrony achieved between the two subsystems. From equations (\ref{cv}) and (\ref{cv2}) we found $\frac{c}{c^\prime}\sim\frac{x_\alpha+x_\beta}{x_\alpha}~\rangle 1$ giving rise $c\rangle c^\prime$. Again from equation (\ref{cv2}) and (\ref{cvm}), we could arrive at $\frac{c^\prime}{\epsilon}\sim\frac{\eta}{x_\alpha+x_\beta}\langle 1$ with $B\sim D$ such that $c^\prime\langle\epsilon$. Similarly, from equation (\ref{cvm}), (\ref{cvm2}), (\ref{cv2}) and (\ref{cvm}), we can show that $\frac{\epsilon}{\epsilon^\prime}\sim\frac{\eta+\psi}{\eta}\rangle 1$ and $\frac{c^\prime}{\epsilon^\prime}\sim\frac{\eta+\psi}{x_\alpha+x_\beta}\rangle 1$ respectively giving rise $\epsilon\rangle\epsilon^\prime$ and $c^\prime\ge\epsilon^\prime$. Thus summarizing the possible relations, we have,
\begin{eqnarray}
\label{comp}
c~\ge~\epsilon~\rangle~ c^\prime~\ge~\epsilon^\prime
\end{eqnarray}
This indicates that the synchronization of the oscillators occurs at largest value of coupling constant in the case of single molecule diffusive coupling and at the smallest value of it in the case of double molecule mean-field coupling. In another words, the rate of information processing from one oscillator to another is fastest in the case of double molecule mean-field coupling mechanism and the rate of this information processing in ascending order is given by equation (\ref{comp}).
\begin{figure*}
\label{fig4}
\begin{center}
\includegraphics[height=300 pt]{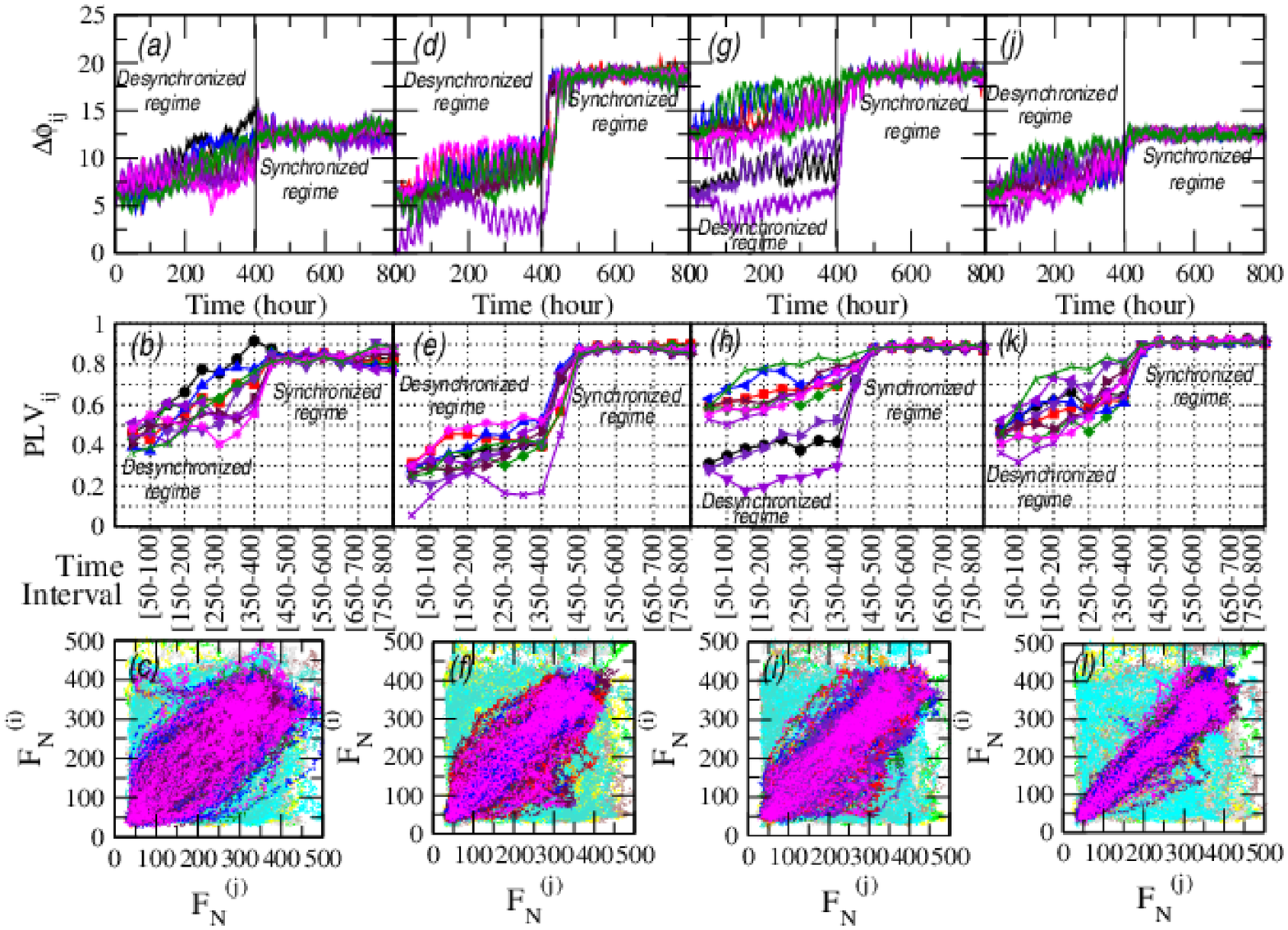}
\caption{Plots showing transition from desynchronized to synchronized regime for different coupling mechanisms at $V=100$ and $c(c~or~c^\prime~or~\epsilon,~or~\epsilon^\prime)=0.6$: (1) Diffusive coupling (single molecule): $M$ is taken as coupling molecule. Phase plots, phase locking values ($PLV$) and recurrence plots are shown in panels (a), (b) and (c). (2) Diffusive coupling (double molecule): $M$ and $F_C$ are taken to coupling molecules. The respective plots are as in (1) are shown in panels (d), (e) and (f). (3) Mean-Field coupling (single molecule): $M$ is taken as coupling molecule and mean field for 30 oscillators is taken. The respective plots as in (1) are shown in panels (g), (h) and (i). (4) Mean-Field coupling (double molecule): $M$ and $F_C$ are taken as coupling molecules and the corresponding plots are shown in panels (j), (k) and (l).}
\end{center}
\end{figure*}

\begin{figure*}
\label{fig5}
\begin{center}
\includegraphics[height=300 pt]{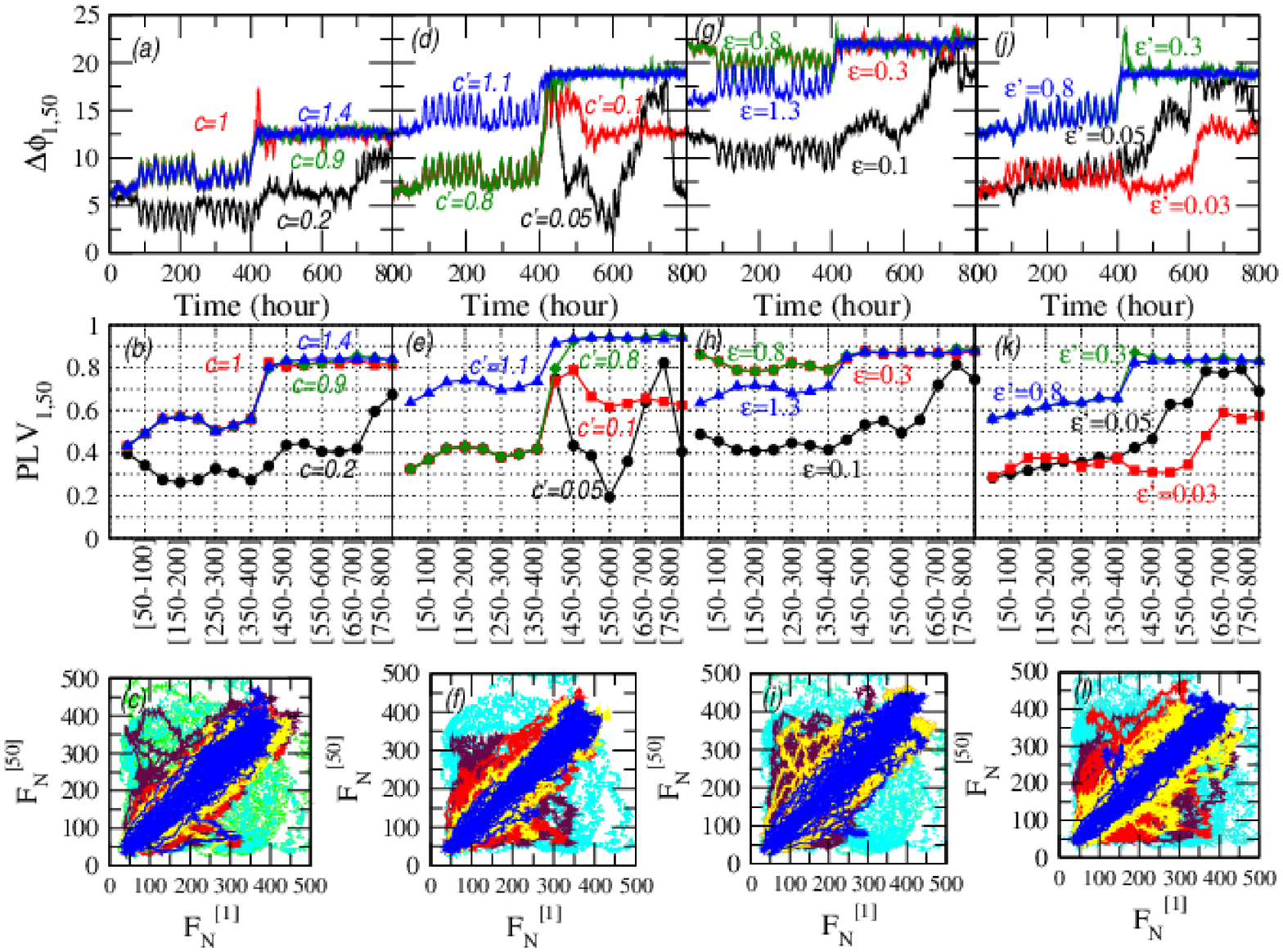}
\caption{ Similar plots as in Fig. 4 are shown but as a function of coupling parameters; Nearest neighbour (1) Diffusive coupling (single molecule) in (a), (b) and (c); (2) Diffusive coupling (double molecule) in (d), (e) and (f); (3) Mean-Field coupling (single molecule) in (g), (h) and (i); (4) Mean-Field coupling (double molecule) in (j), (k) and (l).}
\end{center}
\end{figure*}

\subsubsection*{Measurement of rate of synchrony}

It has been pointed out that the identification of phase synchrony of any two identical systems can be done qualitatively by the measurement of the time evolution of instantaneous phase difference of the two systems \cite{sak,pik,ros,nan}. It is possible if one defines an instantaneous phase of an {\it arbitrary} signal $x(t)$ via Hilbert transform \cite{ros}
\begin{eqnarray}
\label{phase}
\tilde x(t)=\frac{1}{\pi}P. V. \int_{-\infty}^{+\infty}\frac{x(t)}{t-\tau}d\tau
\end{eqnarray}
where $P. V.$ denotes the Cauchy principal value. Then, one can determine an instantaneous "phase" $\phi(t)$ and an instantaneous "amplitude" $A(t)$ of the given signal through the relation, $A(t)e^{i\phi(t)}= x(t)+i\tilde x(t)$. Given two signals of two systems $x_i(t)$ and $x_j(t)$, one can therefore obtain instantaneous phases $\phi_i(t)$ and $\phi_j(t)$; phase synchronization is then the condition that $\Delta\phi_{ij}=m\phi_i-n\phi_j$ is constant, where $m$ and $n$ are integers. Of most interest are the cases $\Delta \phi_{ij}$ = 0 or $\pi$, namely the cases of in--phase or anti--phase, but other temporal arrangements may also occur. 

The phase synchronization of the two identical systems can also be identified by doing synchronization manifold recurrence plot of the variable of one system, say $x$ with the corresponding variable $x^\prime$ of the other system simultaneously on two dimensional cartesian co-ordinate system \cite{pec}. The rate of synchronization between the two systems can be determined by the rate of concentration of the points in the plot along the diagonal. If the two systems are uncoupled, then the points on the plot will scatter away randomly from the diagonal.

The rate of phase synchrony of the two systems can be estimated quantitatively by measuring the "phase locking value" (PLV) of the two signals of the two systems \cite{lac}. The phase locking value, which is used to quantify the degree of synchrony, characterizes the stability of phase differences between the phases $\phi_i(t)$ and $\phi_j(t)$ of two signals $x_i(t)$ and $x_j(t)$ of ith and jth systems and can be defined within a time period $T$ by,
\begin{eqnarray}
\label{plv}
P(t)=\frac{1}{T}\left|\sum_{t-T}^t\Delta\phi_{ij}\right|
\end{eqnarray}
The range of PLV value is [0,1]. When the value of PLV is zero, the two systems are uncoupled, whereas if the value of PLV is one then the two systems are perfectly phase synchronized \cite{lac}.

\subsection*{Application: N. crassa circadian model}

The circadian model we consider is the simplified reduced model \cite{gon1,gon2,gon3,lel1,lel2} in which the negative feedback mechanism is incorporated during genetic regulation of the clock protein \cite{gol} as shown in Fig.1. We briefly describe the reaction mechanisms of the model as follows. The biochemical network of the circadian rhythm involve transcription and transportation of mRNA ($M$) with rate $\nu_s$ by clock gene ($frq$) into cytosol. Then $M$ decays with rate $\nu_m$ or transported into cytosolic protein ($F_C$) with rate $k_s$. $F_C$ either decays with rate $\nu_d$ or get inside the nucleus to form $F_N$ with rate $k_1$ which is a reversible reaction. This gives rise six reaction steps: $\phi\stackrel{\nu_s}\rightarrow M$, $M\stackrel{\nu_m}\rightarrow\phi$, $\phi\stackrel{k_s}\rightarrow F_C$, $F_C\stackrel{\nu_d}\rightarrow\phi$, $F_C\stackrel{k_1}\rightarrow F_N$, $F_N\stackrel{k_2}\rightarrow F_C$; with transition rates given by, $\chi_1=\nu_sV\frac{[K_IV]^n}{[K_IV]^n+F_N^n}$, $\chi_2=\nu_mV\frac{M}{K_mV+M}$, $\chi_3=k_sM$, $\chi_4=\nu_dV\frac{F_C}{K_dV+F_C}$, $\chi_5=k_1F_C$ and $\chi_6=k_2F_N$. The reactions are derived from the set of classical differential equations, which describe the molecular mechanisms, to stochastic reaction steps using transformation equation, $\Gamma(X)=V\gamma_i(\frac{X}{V})$, where $\Gamma_i$ and $\gamma_i$ are stochastic and classical transition rates, $X=[M,F_C,F_N]^{-1}$ and $x=\frac{X}{V}=[m,f_c,f_N]^{-1}$ are state vector of the molecular populations and concentrations, and $V$ is dimensionless system size.

In the reduced reaction network model of N. crassa which exhibit circadian rhythm, the stochastic state of the system at any instant of time $t$ can be described by a state vector $X(t)=[M(t),F_C(t),F_N(t)]^{-1}$. At any stochastic state of the system, the participating molecules in the network suffer decay or creation of molecules \cite{gil1}. The time evolution of the transitional probability of the stochastic states is described by a Master equation \cite{mcq,gil1}. The stochastic dynamics was simplified by Gillespie based on identification of reaction number and reaction time, and we use this algorithm to simulate time evolution of molecular populations. 
\begin{figure}
\label{fig6}
\begin{center}
\includegraphics[height=160 pt]{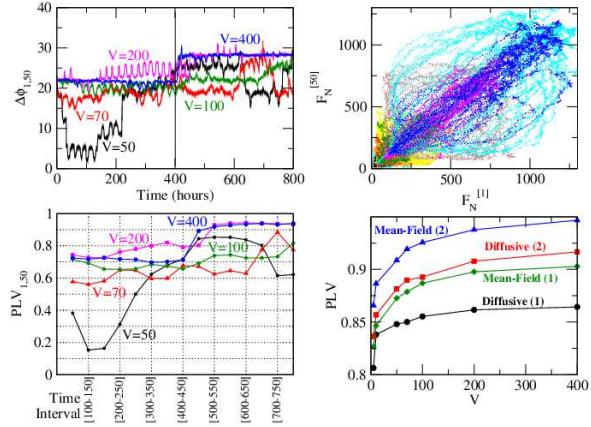}
\caption{Phase plots ($\Delta\phi$ vs $t$), $PLV$ and recurrence plots for various values of $V$ for diffusive coupling (single molecule). The lower right hand panel shows $PLV$ as a function of $V$ for all four types of coupling mechanisms.}
\end{center}
\end{figure}

In our numerical simulation, we consider two dimensional array of oscillators $(N\times L)$ coupled by nearest neighbour diffusive or mean-field coupling as shown in Fig 2. We considered fixed boundary conditions i.e. for molecular species $F_N$; $F_N^{[N+1,L]}=0$, $F_N^{[0,L]}=0$, $F_N^{[N,L+1]}=0$, $F_N^{[N,0]}=0$ and similarly the same condition is applied for the remaining other molecular species. So for any oscillator, there are at least two and at most four nearest neighbour coupled oscillators.

\subsection*{Results and discussions}

The number of nuclei which can be viewed as self-sustained oscillators present in a single cytoplasm of N. crassa is small and finite. The oscillators are coupled via various coupling mechanisms mentioned in the previous section and the oscillators naturally prefer to choose the coupling mechanism that enable to process information quicker and easier i.e. in another word the mechanism which enable the oscillators synchronize fastest. In our large scale simulation we use stochastic simulation algorithm due to Gillespie \cite{gil1} which we developed in Java language and we take 50 oscillators (nuclei) in a N. crassa single cell. During our simulation we also assume that these nuclei are static relative to each other because of the reasons that the rate of diffusion of the diffusing molecules or proteins ($M$, $F_C$ etc) from one oscillator to another is much much faster as compare to the oscillator motion and there are other cellular processes such as microtubule, spindle etc that cause resistance in their relative motion. 

We first present the results of nearest neighbour single molecule bidirectional diffusive coupling among the 50 oscillators where $M$ is taken to be coupling molecule as shown in Fig. 3. The simulation is done at system size $V=100$ with coupling constant $c=1.4$ and coupling is switch on at $time=400 hours$. The upper two panels show the $F_N$ dynamics as a function of time for 10 oscillators and their corresponding phase difference $\Delta\phi$ for each pairs of oscillators as a function of time calculated using Hilbert transform (\ref{phase}). When the oscillators are uncoupled, the $F_N$ dynamics of the oscillators and their corresponding $\Delta\phi$ are evolved independently of each other, whereas when the oscillators are coupled, the $F_N$ dynamics exhibit the same correlated behaviour and their corresponding $\Delta\phi$ fluctuate about a constant value separating synchronized and desynchronized regimes. This phase transition like behaviour separating synchronized and desynchronized regimes is again verified by lowermost two panels the phase locking values, $PLV$ dynamics of the oscillators and thier synchronization manifold recurrence plot. In desynchronized regime, the $PLV$ evolved independently and in synchronized regime it remain at a constant value near the value 1. In the case of synchronized manifold recurrence plot, the synchronized regime the points are concentrated along the diagonal, whereas in the desynchronized regime, the points spread away from the diagonal.

Now we compare the rate of synchrony for four different coupling mechanisms at constant system size, $V=100$ and coupling constant $c=0.7$ in Fig. 4. The results due to one molecular species diffusive coupling (coupling molecule is $M$) are shown in Fig. 2 such that (a) phase plot i.e. $\Delta\phi$ vs $t$, (b) phase locking value plot i.e. $PLV$ vs $t$ and (c) synchronized manifold i.e. recurrence plot. The plots show that the rate of synchrony is very weak because of the large fluctuation in the $\Delta\phi$ and $PLV$ vs $t$ plots in synchronized regime as well as larger spreading of points away from the diagonal. Similarly, the results for double molecular species diffusive coupling, single and double molecular species mean-field coupling mechanisms are shown in Fig. 4 (d), (e), (f); Fig. 4 (g), (h), (i); and Fig. 4 (j), (k), (l) respectively. The plots in all cases of coupling mechanisms show that the rate of synchrony is strongest in the case of double molecular species mean-field coupling and weakest in the case of single molecular species diffusive coupling. Our simulation results support our theoretical claim in the expression (\ref{comp}).

Next we present our simulation results for the four coupling mechanisms at constant system size $V=100$ and for different values of coupling constants as shown in Fig. 3. The results of single molecular species diffusive coupling are shown in Fig. 5 (a), (b) and (c) respectively. It shows that for lower values of $c$, the $\Delta\phi$ and $PLV$ evolve randomly as the function of time even if the coupling is switch on as shown in Fig. 5 (a) and (b). However this random fluctuation starts co-ordinating around a constant value as $c$ increases and we found the rate of synchrony to be strongest at $c=1.4$ and remain the same rate for $c\ge1.4$. This claim is supported by our recurrence plot in Fig. 5 (c). The results of double molecular species diffusive coupling, single and double molecular species mean-field coupling mechanisms are shown in Fig. 5 (d), (e), (f); Fig. 5 (g), (h), (i); and Fig. 5 (j), (k), (l) respectively. The values of coupling constants at which rate of synchrony is strongest in these three coupling mechanisms are found to be $c=1.1,~1.3$ and $0.8$ respectively.

We show our simulation results for the single molecular species diffusive coupling mechanism at constant coupling constant $c=0.6$ and for different values of system sizes in three panels of Fig. 6. The results show that even if the coupling is switch on at $t=400$ hours, the dynamics of $\Delta\phi$ and $PLV$ randomly evolve with as a function of $t$ for system sizes $V\langle400$. However the rate of synchrony is strongest at $V=400$ and remain the same rate for system sizes $V\ge400$. Our claim is again supported by the recurrence plot shown in third panel of Fig. 6. Then we compare the $PLV$ values for different values of $V$ for four different coupling mechanisms as shown in fourth panel of Fig. 6. The plots show that the rate of synchrony is strongest for double molecular species mean-field coupling scheme and weakest for single molecular species diffusive coupling scheme again supporting our theoretical claim (\ref{comp}).

We finally compare the four coupling mechanisms in the phase diagram shown in Fig. 7 separating desynchronized and synchronized regime. The plot clearly indicates that the rate of synchrony in double molecular species mean-field coupling scheme is strongest and quickest since the area bounded by the curve in synchrony regime is largest. Whereas the rate of synchrony in single molecular species diffusive coupling scheme is weakest and slowest since the area bounded by the curve in synchrony regime is least. This clearly support our theoretical claim in (\ref{comp}) again.
\begin{figure}
\label{fig7}
\begin{center}
\includegraphics[height=110 pt]{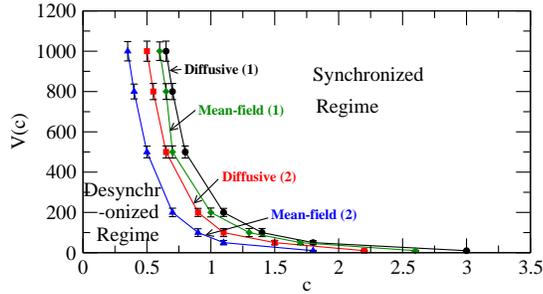}
\caption{Plot of phase diagram in which $V$ for all four coupling mechanisms are shown as a function of their corresponding coupling constants $c$ ($c$ could be $c$, $c^\prime$, $\epsilon$ or $\epsilon^\prime$ according to their respective coupling mechanisms) separating synchronized and desynchronized regimes.}
\end{center}
\end{figure}

\subsection*{Conclusion}

Naturally the coupled oscillators process information from one oscillator to another by selecting the coupling mechanism which gives fastest information processing and easily available in nature out of different coupling schemes. This idea of selection of coupling mechanisms by the coupled oscillators can be seen by measuring the rate of synchrony where this rate is maximum. In such situation of maximum rate of synchrony, the rate of information processing among the coupled oscillators is quickest and probably the agents which are responsible for information transfer among the oscillators may be easily available.

In our large scale simulation for different coupling mechanisms that might happen in real situation of N. crassa circadian model, we found that mean-field coupling mechanism might be a prefered coupling scheme out of four schemes we studied. This numerical results support our theoretical claim also. The coupling schemes we studied so far allow relay or long range information transfer among the coupled oscillators. However there are different coupling mechanisms such as direct coupling, delay time coupling etc which we do not study here thinking that either these coupling mechanisms are not realistic or slow coupling mechanisms specially in our N. crassa system.

\subsection*{Acknowledgments}
This work is financially supported by Department of Science and Technology (DST), New Delhi under sanction no. SB/S2/HEP-034/2012.

\end{document}